\documentclass[preprint,3p,times,twocolumn]{elsarticle}

\usepackage{xcolor}
\usepackage{soul}
\usepackage{amsmath}

\usepackage{amssymb}
\usepackage{subfig}

\journal{Spectrochimica Acta Part B}

\begin{document}

\begin{frontmatter}



\title{Absolute frequency measurement of the 5s5p $^1P_1$ - 5s5d $^1D_2$ transition in strontium}


\affiliation[IFS]{organization={Institute of Physics, Centre for Advanced Laser Techniques},
            addressline={Bijenička cesta 46}, 
            city={Zagreb},
            postcode={10000}, 
            country={Croatia}}

\author[IFS]{Ana Cipriš\fnref{fn1,fn2}}
\author[IFS]{Ivana Puljić\fnref{fn1}}
\author[IFS]{Damir Aumiler}
\author[IFS]{Ticijana Ban}
\author[IFS]{Neven Šantić\corref{cor1}}

\fntext[fn1]{These authors contributed equally to this work.}
\fntext[fn2]{Present address: Instituto de Física de São Carlos, Universidade \hspace*{1.5em} de São Paulo, São Carlos, SP 13566-970, Brazil}
\cortext[cor1]{Corresponding author: nsantic@ifs.hr}

\begin{abstract}
We report on the absolute frequency determination of the 5s5p $^1P_1$ - 5s5d $^1D_2$ transition in atomic strontium, achieved through frequency comb-referenced laser-induced-fluorescence (LIF) spectroscopy. 
We excite the 5s$^2$ $^1S_0$ - 5s5p $^1P_1$ transition using an on-resonance laser at $\approx$461 nm, and then measure the variation in the LIF signal while scanning the laser at $\approx$767 nm across the 5s5p $^1P_1$ - 5s5d $^1D_2$ transition.  
We determine the absolute frequency of $390599571.7\pm0.4$ MHz, with an accuracy that surpasses the previous most accurate measurement by two orders of magnitude. 
This measurement technique can be readily applied for precision spectroscopy of high-lying states not only in strontium, but also in other atomic species.

\end{abstract}




\end{frontmatter}

\section{Introduction}
\label{sec:intro}
In the last decade, strontium has emerged as a key atomic species in a wide array of cutting-edge cold-atom-based quantum technologies. These technologies encompass optical atomic clocks \cite{Ludlow2015} which have achieved record precision \cite{Marti2018,Bothwell2022}, single-photon atomic interferometers \cite{Hu2017,Hu2019,Rudolph2020}, tweezer-based quantum simulators \cite{Kaufman2021,Cooper2018,Levine2018}, and neutral-atom quantum computers \cite{Schine2022,Pagano2022}. 
The favorable electronic levels structure of atomic strontium is at the heart of these applications.
For example, strontium $^3P_0$ - $^1S_0$ clock transition features the longest-lived optical excited electronic state in a neutral atom \cite{Muniz2021}, a key feature for state-of-the-art atomic clocks and quantum computing.  
Furthermore, strontium offers two transitions for laser cooling, a strong $^1S_0$~-~$^1P_1$ transition at 461 nm with a natural linewidth of 2$\pi$ x 32~MHz, and a weak $^1S_0$~-~$^3P_1$ transition at 689 nm with a natural linewidth of 2$\pi$ x 7.5~kHz \cite{Sansonetti2010}.
Both of these transitions can be driven using standard semiconductor laser diodes. 
Notably, the transition at 689 nm, with a Doppler temperature of merely 179~nK, allows for direct laser cooling to quantum degeneracy \cite{Stellmer2013}.

A deeper understanding of the electronic structure of atomic strontium, including its higher-lying electronic states, is essential for enhancing the performance of all the aforementioned applications. 
A prime example is the case of the strontium optical clock, where accuracy depends on our understanding of the black body radiation shift \cite{Safronova2013,Middelmann2012,Nicholson2015,Hobson2020,Lisdat2021}. 
Correcting this shift relies on theoretical models of the atom's internal structure, which, in turn, depend on precise spectroscopic measurements of the various electronic states in the strontium atom. 
Furthermore, accurate spectroscopic measurements of optical transitions between higher electronic states are valuable for generating strontium Rydberg states that exhibit rich many-body quantum behavior and are used to realize fast two-qubit quantum logic gates \cite{Hollerith2022}.

Spectroscopic measurements of optical transitions in strontium using Fabry-Pérot interferometry \cite{Sullivan1938}, arc emission spectroscopy \cite{Garton1968}, and laser induced fluorescence \cite{Saunders1910} can be found in literature.
However, their accuracy is limited, highlighting the need for further measurements to enhance precision and accuracy.

In this work, we present a measurement of the absolute frequency of the 5s5p $^1P_1$ - 5s5d $^1D_2$ transition in atomic strontium. 
A strontium atomic beam is generated in an under-vacuum spectroscopy cell.
We first populate the 5s5p $^1P_1$ state using an on-resonance diode laser at $\approx$461 nm.
Subsequently, we probe the 5s5p $^1P_1$ - 5s5d $^1D_2$ transition with a continuous-wave (cw) titanium-sapphire (Ti:Sa) laser at $\approx$767 nm.
The laser-induced fluorescence (LIF) is measured as a function of the Ti:Sa laser frequency.
Simultaneously with the LIF signal, we detect a beat signal between the Ti:Sa laser and an optical frequency comb.
The frequency comb (FC) is fully stabilized and referenced to a primary frequency standard, thus providing the absolute frequency scale for LIF measurements.
The measured 5s5p $^1P_1$ - 5s5d $^1D_2$ transition frequency is $390599571.7\pm0.4$ MHz, with an accuracy that surpasses the previous most accurate measurement \cite{Sullivan1938} by two orders of magnitude.
Details of the experiment are given in Section \ref{sec:exp}, our results are presented in Section \ref{sec:results}, and we discuss all systematic effects contributing to the error budget in Section \ref{sec:discussion}. 
We conclude in Section \ref{sec:conclusion}.

\begin{figure}[h]
\centering
\includegraphics{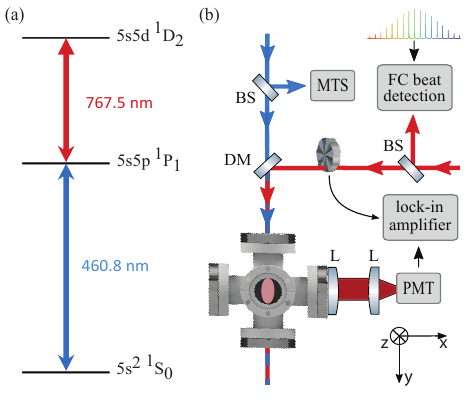}
\caption{(a) The relevant energy levels of strontium atom. (b) A simplified scheme of the experimental setup. MTS - modulation transfer spectroscopy, DM - dichroic mirror, BS - beamsplitter, PMT - photomultiplier tube, L - lens, FC - frequency comb.}
\label{fig:scheme}
\end{figure}

\section{Experimental setup}
\label{sec:exp}

A simplified scheme of the experimental setup is shown in Fig. \ref{fig:scheme}(b), with the relevant energy levels of strontium atom shown in Fig. \ref{fig:scheme}(a).
We utilize an under-vacuum spectroscopy cell to generate an atomic beam of hot strontium atoms using dispensers (AlfaVakuo, AS-Sr-5V-600).
The cell design closely resembles that described by Bridge et al.~\cite{Bridge2009}, featuring an additional viewport for collecting LIF perpendicular to the direction of the laser beams and the atomic beam.

We first populate the $^1P_1$ state using an on-resonance cw external cavity diode laser (MOGLabs, CEL) at $\approx$461 nm (hereinafter referred to as the blue laser).
Subsequently, we probe the $^1P_1$ - $^1D_2$ transition with a cw Ti:sapphire laser (Sirah Lasertechnik) at $\approx$767 nm (hereinafter red laser).

Both blue and red lasers have the same linear polarization.
The laser beams overlap along the entire spectroscopy cell, propagating through the centre of the cell in a co-propagating or counter-propagating geometry.  
Unless specified differently, the powers of the blue and red lasers (measured at the entrance of the cell) were 250~$\mu$W and 150~$\mu$W, with $1/e^2$ waists of 1.2~mm and 0.96~mm, respectively.
This corresponds to the $I_{\text{blue}}/I_{\text{sblue}}$=0.3 and $I_{\text{red}}/I_{\text{sred}}$=26, where $I_{\text{blue}}$ and $I_{\text{red}}$ are corresponding laser intensities, while $I_{\text{sblue}}$=42~mW/cm$^2$ and $I_{\text{sred}}$=0.4~mW/cm$^2$ are the saturation intensities of the $^1S_0$ - $^1P_1$ and $^1P_1$ - $^1D_2$ transitions calculated from \cite{Sansonetti2010, Werij1992}.

The frequency of the blue laser is stabilized using modulation transfer spectroscopy (MTS) \cite{McCarron2008}, in another under-vacuum spectroscopy cell which is identical to the main cell.
The frequency of the red laser is stabilized to a wavemeter (HighFinesse, WS7-100) with a digital PID loop feeding back on the slow in-cavity piezo actuator. 

The frequency of the red laser is measured using an optical frequency comb ~\cite{Ye2005}.
The frequency comb (MenloSystems, FC1500-ULN), with a repetition frequency of 250 MHz, is referenced to a 10 MHz primary frequency standard (Microchip, 5071B cesium atomic clock).
The frequency of the red laser, $f$, is determined using the relation:
\begin{equation}\label{eq:beat}
    f = \textit{N}f_{\text{rep}} \pm 2f_{\text{ceo}} \pm f_{\text{beat}},
\end{equation}
where $\textit{N}$ is the nearest adjacent comb mode number, $f_{\text{rep}}$ is the frequency comb repetition rate, $f_{\text{ceo}}$ is the frequency comb carrier–envelope offset frequency, and $f_{\text{beat}}$ is the measured beat note between the red laser and the nearest adjacent comb mode with the mode number $\textit{N}$. 
The $f_{\text{rep}}$ and $f_{\text{ceo}}$ are measured using a fast photodiode and f-2f interferometer, respectively, and frequency stabilized via phase lock to the RF standards.
$N$ is unambiguously determined by measuring the frequency of the red laser using a precise wavemeter (HighFinesse, WS7-100) which has an accuracy of 100 MHz, i.e. the wavemeter accuracy is better than half the
repetition rate.
The signs of $f_{\text{ceo}}$ and $f_{\text{beat}}$ are determined by changing $f_{\text{ceo}}$ and $f_{\text{rep}}$ both in locked condition while monitoring the resulting change of the beat note $f_{\text{beat}}$.
The beat note signal, $f_{\text{beat}}$, is measured with a fast photodetector (Menlo Systems APD210) followed by a high-resolution multichannel frequency counter (K+K Messtechnik, FXE) with a gate time of 1 s. 
In our experiment, $f_{\text{rep}}\approx$250~MHz, $f_{\text{ceo}}=35$ MHz, and N=1562395. 

The LIF (schematically shown as a pink area in Fig. \ref{fig:scheme}(b)) is detected using a photo-multiplier tube (PMT, Hammamatsu, H6780-02). 
Using a telescope with a 5:1 focal ratio, the whole $\approx40$~mm fluorescence stripe was imaged onto the 8~mm effective area of the photo-multiplier tube sensor. 
The LIF signal comprises two contributions stemming from the "blue" $^1P_1$ - $^1S_0$ and "red" $^1D_2$ - $^1P_1$ transitions, respectively.
The dominant contribution to the LIF signal arises from the "blue" $^1P_1$ - $^1S_0$ transition, given that it is the strongest transition in strontium with a scattering rate of $\Gamma_{\text{blue}}$=2$\pi$ x 32~MHz \cite{Sansonetti2010}. 
For comparison, the scattering rate of the "red" $^1D_2$ - $^1P_1$ transition is $\Gamma_{\text{red}}$=2$\pi$ x 1.5~MHz \cite{Sansonetti2010}. 
As the frequency of the red laser is scanned across the $^1P_1$ - $^1D_2$ resonance, a fraction of the atomic population is transferred from the $^1P_1$ to the $^1D_2$ state.
Consequently, as we approach the $^1P_1$ - $^1D_2$ resonance, the red fluorescence intensity increases while the blue fluorescence decreases.
To detect only the blue or red fluorescence we use either a shortpass (Thorlabs, FESH0500) or a longpass (Thorlabs, FELH0500) filter, respectively.
Other than the opposite change in intensity, the LIF spectra obtained from collecting both "blue" and "red" fluorescence exhibit identical behavior. 
Nevertheless, owing to the superior signal-to-noise ratio of the "blue" LIF, all measurements presented in this paper were performed using a shortpass filter, i.e. detecting the blue fluorescence.
Additionally, to measure only the change in the "blue" LIF  resulting from the excitation induced by the red laser on the $^1P_1$ - $^1D_2$ transition, the standard lock-in technique is employed.  
The red laser is modulated at 8 kHz using a mechanical chopper (Thorlabs, MC1F10HP), and the signal from the photo-multiplier tube is fed into a lock-in amplifier (SRS, SR830).
 
To probe the $^1P_1$ - $^1D_2$ transition, we scan the red laser in steps of $\sim$10 MHz over the range of up to several hundreds of MHz.
For each frequency of the red laser (measured with a wavemeter), we simultaneously measure the LIF signal and the frequency of the optical beat. 
We average both signals, i.e. LIF and beat, over typically 100 s.
We determine the absolute frequency of the red laser using the Eq. \ref{eq:beat}, and in this way we obtain one point in the LIF spectrum.
In order to reconstruct the whole $^1P_1$ - $^1D_2$ transition lineshape, around 30 points in LIF spectrum are measured.

\section{Absolute frequency of the $^1P_1$-$^1D_2$ transition in strontium}
\label{sec:results}

For precise determination of the $^1P_1$-$^1D_2$ transition frequency from the measured LIF spectra, meticulous consideration of the detuning of the blue laser from the $^1S_0$-$^1P_1$ resonance is essential.
This detuning plays a critical role in influencing the observed spectral line.
Indeed, an unknown detuning, $\delta$, of the blue laser from the $^1S_0$-$^1P_1$ resonance, subjected to the accuracy of our locking technique, results in the observation of a frequency-shifted $^1P_1$-$^1D_2$ spectral line.
It is determined by $\pm$ $\delta \cdot k_2/k_1$, where $k_1$ and $k_2$ denote the wave vectors of the blue ($^1S_0$-$^1P_1$ transition) and red ($^1P_1$-$^1D_2$ transition) lasers, respectively.
The frequency shift has opposite signs for lasers propagating in counter- and co- propagating geometries.
The unshifted, i.e., the absolute, frequency of the $^1P_1$-$^1D_2$ transition lies precisely midway between the peak frequencies of the spectral lines in the co- and counter-propagating lasers geometries. 
For a more comprehensive understanding of the influence of the blue laser detuning on the determination of the  $^1P_1$-$^1D_2$ frequency see the supplemental materials, where we present the results of the theoretical simulations based on the Optical Bloch Equations (OBE).

To determine the absolute frequency, $f$, of the $^1P_1$-$^1D_2$ transition, we measured LIF spectra using both co- and counter-propagating geometries of the red and blue lasers for several blue laser detunings.  
For each detuning of the blue laser we determine the peak frequencies of the spectral lines using co- and counter-propagating geometries, $f_{\text{co}}$ and $f_{\text{counter}}$, respectively, from which the absolute frequency of the $^1P_1$-$^1D_2$ transition is calculated as $f=(f_{\text{co}}+f_{\text{counter}})/2$.

\begin{figure}[h]
\centering
\includegraphics{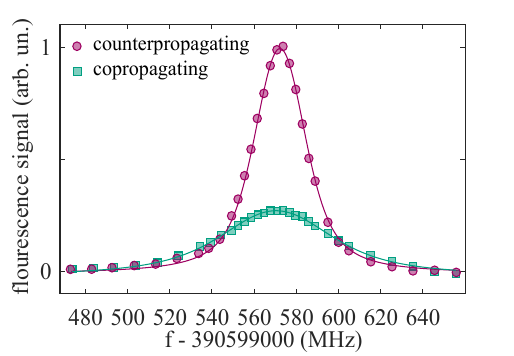}
\caption{The variation in the blue LIF as a function of the red laser frequency for counter-propagating (red, circles) and co-propagating (green, squares) geometries. Both horizontal and vertical error bars are smaller than the symbols. The lines are the Voigt fit to the measured data.
}
\label{fig:result}
\end{figure}

In Fig. \ref{fig:result} we show measured LIF spectra using counter-propagating (circles) and co-propagating (squares) geometries for $\delta$=1 MHz. 
$f_{\text{co}}$ and $f_{\text{counter}}$ are obtained by fitting Voigt functions (solid lines) to the measured data.
As expected, the LIF spectrum obtained using the co-propagating configuration appears broader, by a factor of around 2.4, and lower, by a factor around 3.4, compared to the spectrum obtained using the counter-propagating configuration.
The measurements are in qualitative agreement with the theoretical results provided in the supplemental materials. 

The results of the absolute frequency determination of the $^1P_1-^1D_2$ transition for three detunings of the blue laser are summarized in Table \ref{table:result_detuning}. 
Each row shows the central frequencies of the spectral lines measured in co- and counter- propagating configurations for a given blue laser detuning.
The last column corresponds to the absolute frequency of the $^1P_1$ - $^1D_2$ transition. 
The blue laser detunings correspond to 0.15, 1, and 5 MHz, respectively for each row of Table \ref{table:result_detuning}. 
These detunings are achieved by introducing a non-zero offset to the error signal generated by the modulation transfer spectroscopy employed to lock the blue laser.
They are calculated from the measured frequency-shifted $^1P_1$-$^1D_2$ spectral lines.
For example, the values shown in the third row of the Table \ref{table:result_detuning} are obtained by shifting the offset by $\approx10$~\% of the peak-to-peak value of the error signal. 
In this specific case, the peak frequencies of the measured spectral lines, using counter- and co-propagating configurations exhibit a difference of 6 MHz, resulting in a frequency shift of 3 MHz. 
Based on this shift, a detuning of the blue laser equal to $\delta$=5 MHz is calculated using the relation $\delta\cdot k_2/k_1$.

\begin{table}[h]
    \begin{center}
 \begin{tabular}{||c|c|c||}
    \hline
    $f_{\text{co}}-f'$ (MHz) & $f_{\text{counter}}-f'$ (MHz)& $f-f'$ (MHz) \\ 
    \hline \hline
    571.7$\pm$ 0.4 & 571.88$\pm$ 0.08 & 571.8$\pm$ 0.2 \\
    571.1$\pm$ 0.2 & 572.35$\pm$ 0.08 & 571.7$\pm$ 0.1\\
    568.7$\pm$ 0.4 & 574.67$\pm$ 0.08 & 571.7$\pm$ 0.2 \\
    \hline
\end{tabular}   
\end{center}
    \caption{The central frequencies of the spectral lines measured in co-, $f_{\text{co}}$, and counter-,$f_{\text{counter}}$, propagating configurations. Each row corresponds to a different detuning of the blue laser from the $^1S_0-^1P_1$ resonance. For clarity, the table displays the deviation from the fixed frequency $f'=390599000$ MHz. The last column is obtained as $f=(f_{\text{co}}+f_{\text{counter}})/2$, and it corresponds to the absolute frequency of the $^1P_1$ - $^1D_2$ transition. The frequency error corresponds to fit errors.  }
    \label{table:result_detuning}
\end{table}

Our measurements give the value of $390599571.7\pm0.1$ MHz for the frequency of the 5s5p $^1P_1$ - 5s5d $^1D_2$ transition.
This value is calculated as a weighted average of the values provided in the third column of Table \ref{table:result_detuning}, with uncertainty that includes both statistical and fit errors.

\section{Sources of uncertainties}
\label{sec:discussion}

In addition to the uncertainty associated with the red laser frequency scale, various sources of error such as Doppler shift, stray magnetic fields, and AC Stark effect contribute to the determination of the overall uncertainty of the $^1P_1$ - $^1D_2$ transition frequency.

\subsection{Frequency scale uncertainty}

The frequency of the red laser is determined using the Eq. \ref{eq:beat}, as described in the Section \ref{sec:exp}.
Consequently, its uncertainty is influenced by the uncertainties associated with determining $f_{\text{rep}}$, $f_{\text{ceo}}$ and $f_{\text{beat}}$.
The accuracies of the $f_{\text{rep}}$ and $f_{\text{ceo}}$ are around 125 $\mu$Hz and 18 $\mu$Hz, respectively, and are determined by the accuracy of the 10 MHz primary frequency standard to which our frequency comb is referenced.
Transferring from the rf to optical domain requires multiplication by N=1562395, resulting in an accuracy of around 200 Hz for $\textit{N}f_{\text{rep}}$.
$f_{beat}$ is measured with a gate time of 1 s and sampling time of 100 s.
The accuracy of the $f_{\text{beat}}$, calculated as the standard deviation of the mean over 100 s of measurements, is 50 kHz. 
This represents the largest uncertainty contribution to the red laser frequency scale.
The corresponding error bars of the frequency scale are too small to be visible in the figures shown.

\subsection{Doppler effect}

When the velocity of atoms in the atomic beam is not perpendicular to the laser beam, a frequency-shifted spectral line is observed.
This shift is known as a Doppler shift.

In our experiment, we utilize two dispensers to generate two atomic beams, each with atomic velocities closely perpendicular to the wave vectors of the blue and red lasers.
To assess the velocity distribution in the ground $^1S_0$ state within this specific geometry, we measured the transmission of the blue laser as a function of its frequency while scanning through the $^1S_0$-$^1P_1$ transition, i.e., we measured an absorption spectrum of the blue laser.
When both dispensers are activated, an atomic velocity distribution that closely resembles the Maxwell-Boltzmann distribution is obtained.
It exhibits a full width at half maximum of $\approx$~360 MHz.

\begin{figure}[h]
\centering
\includegraphics{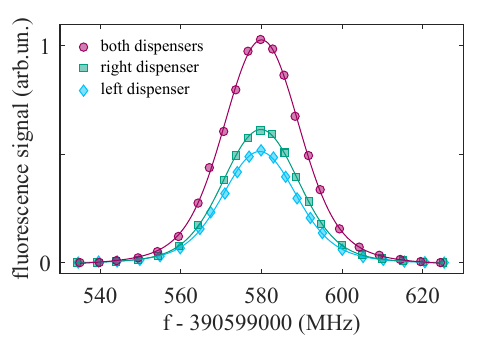}
\caption{The variation in the blue LIF as a function of the red laser frequency for different combinations of the two dispensers available in the experiment. Both horizontal and vertical error bars are smaller than the symbols. The lines are the Voigt fits to the measured data.}
\label{fig:dispensers}
\end{figure}

To evaluate the impact of the Doppler shift on the measured $^1P_1-^1D_2$ transition frequency, we measured LIF spectra as a function of the red laser frequency for various combinations of activated dispensers, as shown in Figure \ref{fig:dispensers}. 
These combinations include both dispensers activated (violet circles), only the right dispenser activated (green circles), and solely the left dispenser activated (blue circles).
The LIF spectra are measured in a counter-propagating configuration of the blue and red lasers.
By fitting Voigt functions (solid lines) to the measured data (symbols), the following peak frequencies are obtained: (580.00$\pm$0.07) for both dispensers activated; (580.1$\pm$0.2) MHz for only right dispenser; and (579.50$\pm$0.08) MHz for only left dispenser.
Based on these values, the uncertainty on frequency measurements caused by Doppler effect is conservatively estimated to $\pm$0.3 MHz. 

\subsection{Stray magnetic fields}

The experiment, not protected by a magnetic shield, is subjected to the earth magnetic field and environmental background magnetic fields, including the field of an ion gauge on the spectroscopy cell.
We measured the total environmental magnetic field to be approximately 3 Gauss, with a longitudinal component along the laser beams propagation axis of $\approx$-1.5~Gauss.

\begin{figure}[h]
\centering
\includegraphics{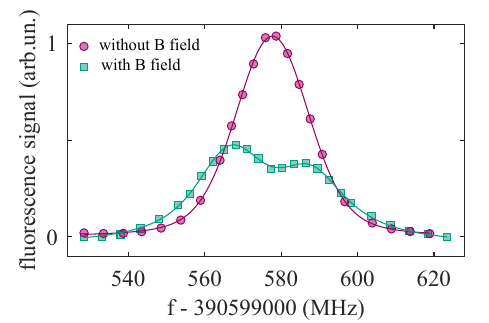}
\caption{The variation in the blue LIF as a function of the red laser frequency measured with (green squares) and without (red circles) an external magnetic field. Both horizontal and vertical error bars are smaller than the point markers. Full lines are Voigt fits to the data.}
\label{fig:magnetic}
\end{figure}

To evaluate the impact of the stray magnetic field on $^1P_1-^1D_2$ frequency measurements an additional DC magnetic field is introduced to the experiment.
The additional magnetic field has a value $\approx$7~Gauss and it is directed along the red laser beam propagation axis, resulting in a total magnetic field of 5.5~Gauss along the laser beams propagation axis.
We then measured the LIF both with and without this additional magnetic field.
The LIF spectra are measured in a counter-propagating configuration of the blue and red lasers.
The blue laser is $\approx$ 10 MHz detuned from the $^1S_0 - ^1P_1$ resonance.
In Fig. \ref{fig:magnetic}, we show the measured spectral lines as a function of the red laser frequency with an additional magnetic field added to the experiment (green squares) and without it (violet circles). 

Under exposure to an additional magnetic field, $^1P_1$ state splits into three components, and $^1D_2$ state splits into five Zeeman components.
The Zeeman splitting is given by $E_{\text{ZE}} = g_{J}\mu_{B}M_{J}B$ and has a value of 1.4 MHz/Gauss for both the $^1P_1$ and $^1D_2$ states \cite{Foot2005}.
The selection rules in our experimental geometry, where the external magnetic field is parallel to the laser propagation vector, are $\Delta M_J = \pm1$ simultaneously.
Consequently, four two-step excitation paths $^1S_0 - ^1P_1 (m_J=-1) - ^1D_2(m_J=-2,0)$ and $^1S_0 - ^1P_1 (m_J=+1) - ^1D_2(m_J=0,+2)$ are possible, resulting in the observation of two Zeeman components in the measured LIF spectrum.
The dominant contribution to the lower frequency Zeeman component arises  from the $^{1}S_0 - ^{1}P_1 (m_J=-1) - ^{1}D_2(m_J=-2)$ transitions.
Likewise, the dominant contribution to the higher frequency component arises from the $^1S_0 - ^1P_1 (m_J=+1) - ^1D_2(m_J=+2)$ transitions.   
When the blue laser is tuned to the $^1S_0 - ^1P_1$ resonance, i.e., for $\delta$=0, the peaks of the two Zeeman components symmetrically split with respect to the LIF measured without a magnetic field.
However, when the blue laser is detuned from the resonance, the splitting of the two Zeeman components becomes asymmetrical in relation to the center of the spectral line measured in the absence of a magnetic field. 
This asymmetry can result in a frequency-shifted LIF spectral line, which can potentially impact the accurate determination of the $^1P_1 - ^1D_2$ absolute frequency.
To estimate this error, we first determine the degree of asymmetric splitting under an additional magnetic field, from the measurement shown in Fig. \ref{fig:magnetic}.
Based on this, we assess the degree of asymmetric splitting occurring due to the influence of stray magnetic fields present in the experiment. This amount is conservatively taken as an error in determining the absolute frequency.

We fit the Voigt function (violet line) to the measured data without an additional magnetic field, Fig. \ref{fig:magnetic}, and obtain a peak frequency of $(577.68 \pm 0.05)$ MHz.  
By fitting a double-peak Voigt function (green solid line) to the data measured with an additional magnetic field, the peak frequencies of the two Zeeman components are determined to be $(566.6 \pm 0.3)$ MHz and $(587.8 \pm 0.4)$ MHz, respectively. 
The separation between the two peaks in the LIF signal is in agreement with the value of the Zeeman splitting for a magnetic field strength of 5.5 Gauss.
The frequency shift of the two components relative to the line measured without an additional magnetic field is 11.1 MHz and 10.1 MHz, respectively.
This suggests an asymmetry of 0.5 MHz when the atoms are exposed to a longitudinal field of 5.5 Gauss, encompassing both the additional and stray magnetic fields. 
Extrapolating to a longitudinal component of the stray field at 1.5 Gauss, an asymmetry of 0.15 MHz is estimated. 
This can be regarded as an error in determining the absolute frequency of the $^1P_1 - ^1D_2$ transition.

\subsection{AC Stark effect}
\label{ACstark}

Another source of systematic error in the determination of the $^1P_1 - ^1D_2$ absolute frequency is the contribution from the AC Stark effect.
As a result of the AC Stark effect, the strongly coupled bare atom energy state splits into two states which are separated by $\hbar\Omega$, where $\Omega$ is the Rabi frequency.
The shift in frequency is evident in the atom's absorption/LIF spectrum, displaying two peaks around the bare transition frequency commonly referred to as Autler-Townes (AT) splitting.

When the blue laser is detuned from the $^1S_0 - ^1P_1$ resonance, the AT splitting of the $^1P_1$ state becomes asymmetric in relation to the bare transition. 
This asymmetry can lead to an observation of frequency-shifted $^1P_1 - ^1D_2$ spectral line, potentially affecting the precise determination of the absolute frequency.
It should be noted that the AT splitting is expected for both the $^1P_1$ and $^1D_2$ states. 
However, considering that the AT splitting is proportional to $\Omega$, the AT splitting of the $^1P_1$ state is always larger than that of the $^1D_2$ state, in our experimental conditions.

\begin{figure}[h]
\centering
\includegraphics{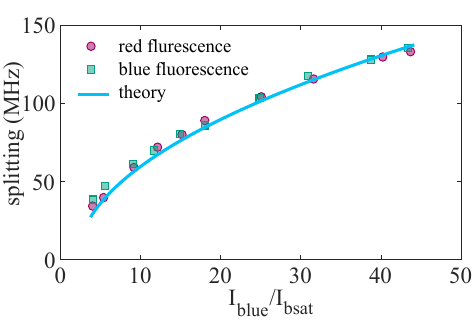}
\caption{Comparison of the calculated (full line) and measured AT splitting as a function of the intensity of the blue laser. Measurements are performed by observing both the red (red circles), and blue (green squares) LIF.}
\label{fig:comparison}
\end{figure}

To evaluate this contribution, we first measured LIF spectra as a function of the frequency of the red laser for several intensities of the blue laser, symbols in Fig. \ref{fig:comparison}.
The intensity of the red laser is maintained at a fixed value of $I_{\text{red}}/I_{\text{sred}}$ = 40.
Here, it's important to emphasize that the $I_{\text{red}}/I_{\text{sred}}$ ratio is deliberately set higher than during absolute frequency measurements from Section \ref{sec:results}, to provide a conservative estimate of the error.
LIF is measured in counter-propagated configuration with linear polarization of the laser fields.
We started the measurements using a high intensity blue laser with $I_{\text{blue}}/I_{\text{sblue}}$=42, for which we observe two fully resolved AT spectral lines.
By fitting a double-peak Voigt function to the measured data, we determined the center frequencies of the split AT components. 
The difference between these frequencies corresponds to the AT splitting for a given intensity of the blue laser.
Then, we gradually decreased the intensity of the blue laser and repeated the procedure.
When the intensity of the blue laser is lowered to $I_{\text{blue}}/I_{\text{sblue}}$=5, the two AT components are no longer resolved.
The AT splitting is obtained from the "blue" (green squares) and "red" (violet circles) LIF spectra, i.e., using a short-pass and a long-pass filter in front of the detector.
As expected, these AT splittings exhibit identical behaviour. 

We then compared our measurements with the results of the theoretical model based on solving the OBE, see Supplementary information for further details.  
The agreement between experiment (symbols) and theory (solid line) is excellent.

Finally, to estimate the error arising from the asymmetric AT splitting, we calculated the frequency shift of the spectral line under conditions corresponding to the experiment used for absolute frequency measurement presented in Section \ref{sec:results}, i.e. $I_{\text{blue}}/I_{\text{sblue}}$=0.3 and $I_{\text{red}}/I_{\text{sred}}$=26.  
Regarding the blue laser detuning, the calculations were performed for the largest detuning of the blue laser used in the $^1P_1 - ^1D_2$ absolute frequency determination, i.e., $\delta=5$ MHz.
The peak of the calculated spectral line is shifted $\approx$ 3.2 MHz from the $^1P_1 - ^1D_2$ resonance. 
As previously explained, a shift of 3 MHz is expected in the counter-propagating configuration for a 5 MHz detuning of the blue laser.
The additional 0.2 MHz shift originates from asymmetric AT splitting. 
Based on this analysis, we conservatively estimated an error in determining the absolute frequency of the 5s5p $^1P_1$ - 5s5d $^1D_2$ transition due to the AC Stark shift to be $\pm$0.2 MHz.

The previously mentioned estimate has also been experimentally validated. 
Under conditions where the AT spectral lines are not yet resolved, i.e. with low intensities of both the blue and red lasers, we measured LIF for different intensities of the blue laser while keeping a constant intensity of the red laser, and conversely, for different intensities of the red laser while keeping the intensity of the blue laser constant.
Each measurement was fitted to a Voigt function, allowing determination of the peak frequency. 
It was observed that all peaks of the measured spectral lines fell within a frequency range of $\pm$0.2 MHz.

\subsection{Overall uncertainty}
\label{Overall}

All pre-mentioned uncertainties have been summarized in Table \ref{table:overall}.
From these results, we deduce an overall uncertainty of $\pm0.4$ MHz.

\begin{table}[h]
    \begin{center}
 \begin{tabular}{||c|c||}
    \hline
    Source of error & Uncertainty (MHz) \\ 
    \hline
    \hline
    Statistical error & $\pm$ 0.1 \\
    Frequency scale determination & $\pm$ 0.05 \\
    Doppler effect & $\pm$ 0.3\\
    Stray magnetic fields & $\pm$ 0.15\\
    AC Stark effect & $\pm$ 0.2\\
    \hline
    overall uncertainty & $\pm$ 0.4\\
    \hline
\end{tabular}   
\end{center}
    \caption{Summary of the estimates on the uncertainties affecting the 5s5p $^1P_1 - 5s5d~^1D_2$ transition frequency measurement.}
    \label{table:overall}
\end{table}

Using these uncertainties, the absolute frequency of the 5s5p $^1P_1 - 5s5d~^1D_2$ transition of strontium is found to be $390599571.7\pm0.4$ MHz.

\section{Conclusion}
\label{sec:conclusion}

In conclusion, we measured the absolute frequency of the 5s5p $^1P_1$ - 5s5d $^1D_2$ transition in strontium. 
The measurement accuracy of the transition frequency is more than two orders of magnitude better than the previous best result \cite{Sullivan1938}. 
Our analysis of the method provides assurance that it can be used for precision spectroscopy of high-lying states in not just strontium, but other atomic species as well. 
This finding could be valuable for future experiments where precise knowledge of transition frequencies in higher-lying electronic states is essential for optimizing the performance of quantum devices.

\section*{Funding}

This work was supported by the following projects: New Imaging and control Solutions for Quantum processors and metrology - NImSoQ, funded through a QuantERA 2021 call; Croatian Quantum Communication Infrastructure – CroQCI, funded through Digital Europe Call (Project Number: 101091513 and NPOO.C3.2.R2-12.01.0001); Frequency comb cooling of atoms - funded through Croatian Science Foundation (IP-2018-01-9047).
In addition, this work was supported by the project Centre for Advanced Laser Techniques (CALT), co-funded by the European Union through the European Regional Development Fund under the Competitiveness and Cohesion Operational Programme (Grant No. KK.01.1.1.05.0001). 

\section*{Declaration of Competing Interest}

The authors declare that they have no known competing financial interests or personal relationships that could have appeared to influence the work reported in this paper.

\section*{Data availability}

Data will be made available on request. 

\section*{CRediT authorship contribution statement}

\textbf{Ana Cipriš:} Investigation, Software, Formal analysis, Writing - Review \& Editing. \textbf{Ivana Puljić:} Investigation, Validation, Formal analysis, Visualization, Writing - Review \& Editing. \textbf{Damir Aumiler:} Software, Validation, Funding acquisition. \textbf{Ticijana Ban:} Conceptualization, Validation, Funding acquisition, Writing - Review \& Editing, Supervision. \textbf{Neven Šantić:} Conceptualization, Validation, Methodology, Writing - Original Draft, Supervision.

 \bibliographystyle{elsarticle-num} 
 \bibliography{cas-refs}





\end{document}